# Experimental demonstration of a time-domain multidimensional quantum channel

**Xingxing Xing[1], Amir Feizpour[1], Alex Hayat[2, 3] and Aephraim M. Steinberg[1, 2]**

[1]*Department of Physics, Centre for Quantum Information and Quantum Control, and Institute for Optical Sciences, University of Toronto, Toronto, Ontario M5S 1A7, Canada*

[2]*Canadian Institute for Advanced Research, Toronto, Ontario M5G 1Z8 Canada*

[3]*Department of Electrical Engineering, Technion, Haifa 32000, Israel*

We present the first experimental realization of a flexible multidimensional quantum channel where the Hilbert space dimensionality can be controlled electronically. Using electro-optical modulators (EOM) and narrow-band optical filters, quantum information is encoded and decoded in the temporal degrees of freedom of photons from a long-coherence-time single-photon source. Our results demonstrate the feasibility of a generic scheme for encoding and transmitting multidimensional quantum information over the existing fiber-optical telecommunications infrastructure.

Multidimensional quantum information (dimension > 2) provides a useful framework for tests of entanglement[1], loophole-free Bell Inequality violation[2] and non-contextuality[3], and higher quantum channel capacity[4]. Unlike photon polarization, external spatial[5] and temporal[6,7,8,9,10,11] degrees of freedom can span in-principle infinite-dimensional Hilbert spaces. Fixed Franson-type[6] interferometers have been employed to demonstrate two-dimensional[12,13] and three-dimensional time-energy entanglement[14]. However, flexible multidimensional quantum information encoding, where Hilbert space dimensionality can be controlled (e.g. by imprinting different sets of phase profiles), has only been demonstrated with photon orbital angular momentum[15,16,17], which are inherently unsuitable for the single-mode fiber-optical communication infrastructure.

We recently proposed[18] a novel flexible multidimensional quantum information scheme based on temporal modulation of single photons, with potential applications in quantum communication and characterization of multidimensional time entanglement. In this time-domain encoding scheme, the multidimensional Hilbert space basis states are an orthonormal, controllable, and (ideally) infinite set of temporal profiles. It was shown theoretically that temporal modulation combined with spectral filtering of photon wavepackets provides a highly efficient encoding technique of controllable-dimension quantum information, compatible with commercially available technology. Single photons produced in a narrow-band source, *e.g.* cavity-enhanced parametric down-conversion (CPDC)[19], can have very long coherence times and be electro-optically (EO) modulated in order to produce a variety of quantum states. In this scheme, the EO modulation prepares the photon in a temporal profile $f_k(t)$ – a "symbol". The photon state can then be described as

$$|\psi_k\rangle = \int d\omega\, \tilde{f}_k(\omega) a_\omega^\dagger |0\rangle \qquad (1)$$

where $\tilde{f}_k(\omega)$ is the Fourier transform of $f_k(t)$. For a projective measurement in a multidimensional Hilbert space, an EOM in a storage loop followed by a spectral filter can be used. In the storage loop, on each round-trip $j$, the photon is modulated by a pattern $f_j^*(t)(f_{j-1}^*(t))^{-1}$, such that after $j$ round trips it has acquired the mode structure $f_j^*(t)$, and the photon state is modified by propagator $U_j$ as

$$U_j|\psi_k\rangle = \int d\omega \left[ \int d\Omega\, \tilde{f}_j^*(\omega) \tilde{f}_k(\omega-\Omega) \right] a_\omega^\dagger |0\rangle.$$ The transmission of this state through a narrow band filter is given by

$$\overline{p}_{kj} \approx T_0^2 \tau_{\text{filter}} \left| \int d\tau f_j^*(t) f_k(t) \right|^2 = T_0^2 \tau_{\text{filter}} \left| \langle \Psi_j | \Psi_k \rangle \right|^2, \qquad (2)$$

which is proportional to the overlap of the *k*-th and *j*-th temporal patterns, with the proportionality constant $T_0^2 \tau_{\text{filter}}$, which is the product of transmission of the filter and its time constant.

We have previously shown theoretically[18] that one can achieve very low error rates based on linear phase modulation, where a linear temporal phase profile is applied to the single-photon pulse, resulting in a frequency shift $\Delta$ in the spectrum,

$$|\psi_k\rangle = \int d\omega\, \tilde{f}_k(\omega-\Delta) a_\omega^\dagger |0\rangle, \qquad (3)$$

If the frequency shift due to the linear phase pattern is large enough so that the overlap between the modulated spectrum and the un-modulated one is negligible, the two states can be considered practically orthogonal. This approach can be used to encode and decode multidimensional quantum information in the frequency basis - the computational

basis - as well as in the basis of frequency superpositions. In this approach, the ability to write and read superpositions of quantum states, which is a crucial property of a quantum communication channel, is provided by the use of EO modulation. Within its modulation bandwidth, the EO modulator allows writing any temporal function on the photon wavepacket; in particular, for any orthogonal basis, $\{|\psi_k\rangle\}$ with a corresponding set of temporal functions, $\{f_k(t)\}$, EO modulation can realize any linear combination of the set $\{f_k(t)\}$, and therefore any superposition of $\{|\psi_k\rangle\}$. In practical EOM-based implementations, the speed of the EOM and the bandwidth of the photon can affect the error rate in the detection[18].

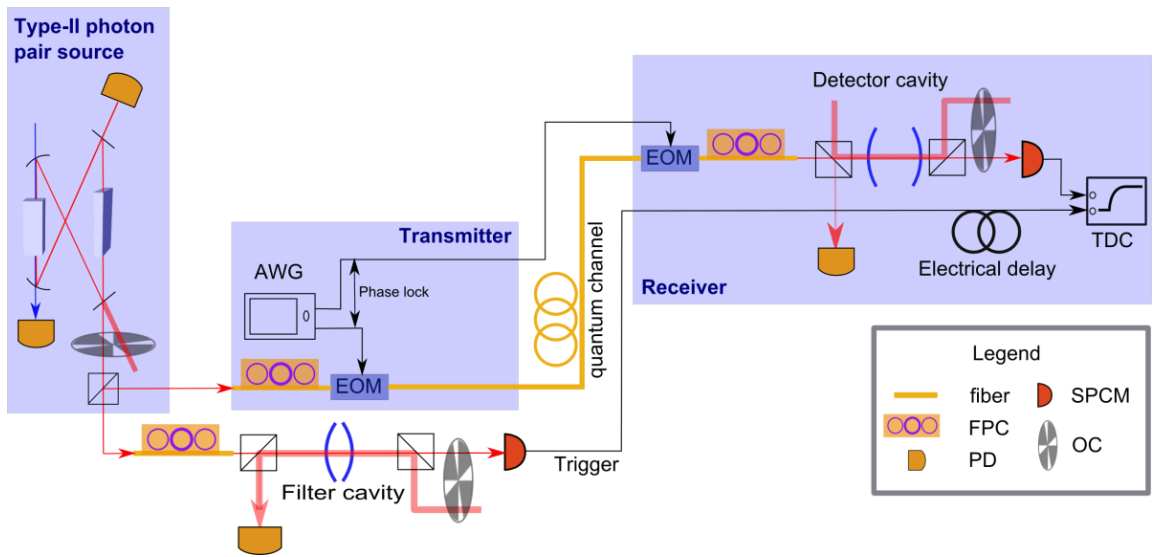

**Figure 1. Experimental setup**. The photon pairs were generated within a PPKTP crystal in a Type-II down-conversion cavity where a KTP crystal was used to compensate the birefringence of the cavity. The idler photon was then sent to a cavity for frequency filtering, which projects the signal photon to a narrow band. The signal photon is then sent through the quantum channel for symbol encoding, transmission, and decoding with EOMs and a detection cavity. After the cavities, the signal and idler are then sent to

SPCMs for subsequent coincidence measurement. FPC: fiber polarization controller, SPCM: single photon counting module, PD: photodiode, OC: optical chopper, AWG: arbitrary waveform generator, TDC: time to digital converter.

In our experimental implementation, in order to encode quantum information in the time domain of a single photon, a single-photon source with long lifetime is required. To achieve this long coherence, we use a heralded single-photon source (Fig. 1) based on cavity-enhanced parametric down-conversion[19] (CPDC). The type II phase-matched nonlinear crystal, PPKTP, was pumped with 20 mW of 390 nm light from a frequency-doubled diode laser. We measured the cavity ring-down time to be $21 \pm 1$ ns, implying a linewidth of $7.6 \pm 0.4$ MHz for each cavity mode. Such a long lifetime enables EO manipulation of the photon wavepacket with commercially available EO modulators. The phase matching bandwidth of the down-conversion is estimated to be 130 GHz at FWHM, much larger than the cavity free spectral range $FSR_{DC}$ of $471 \pm 4$ MHz, thus the cavity output contains many cavity modes. We use a filter cavity to filter out the degenerate modes and reject all others from the multimode emission. To do this, the filter cavity was designed to have a free spectral range larger than the phase matching bandwidth, and a cavity linewidth smaller than $FSR_{DC}$. For the degenerate mode, only one photon from the CPDC pair needs to be filtered, since phase-matching necessarily guarantees the other photon to fall into the same frequency mode. Our filter cavity consists of two curved mirrors (curvature radius of 100 mm) with high reflectivity (R > 99.7%), separated by 1 mm, resulting in a free spectral range $FSR_f =$ 150 GHz. The finesse of the filter cavity is measured to be $329 \pm 5$, and the filter linewidth to be $\delta_f = 455 \pm 46$ MHz.

For such a narrowband (7.6 MHz) source of single photons, frequency stability is crucial to the experiment. This is achieved as follows: a 780 nm external-cavity-stabilized diode laser (Toptica DL100) is locked to the D2 transition of Rb, which is used as an absolute frequency reference for our experiment. The laser light is then amplified and fed to a SHG stage (Toptica TA-SHG-100) to generate the 390nm pump for the CPDC process. The cavity is locked to the frequency-stabilized 780 nm laser using the Pound-Drever-Hall (PDH) technique[20], such that the degenerate mode of the cavity is then also resonant with the frequency reference - the Rb D2 transition. The modulation used in the PDH locking method is at 20 MHz, and the modulation frequency of EOMs is chosen to be 30 MHz to suppress unwanted leakage from PDH modulation. To avoid the difficulty of separating the photon pairs from the classical laser beams used to lock the cavities, we use mechanical choppers with all the cavities to alternate between injection of the locking beam and collection of photon pairs, with the collection duty cycle being 24%. All the mechanical choppers are also phase-locked at a frequency of 200 Hz to synchronize the collection and filtering of the photons.

The encoding and decoding of the information in temporal wavepackets of the single photons (Eq. 3) is done with fast EOMs. The computational basis for our high-dimensional quantum system[18] is defined by a set of discrete frequency bins. The first non-trivial multidimensional quantum system is a qutrit – a three-level quantum system. Our method of encoding quantum information is not limited to qutrits; however, we focus on the three-dimensional Hilbert space of a qutrit to demonstrate the basic concept experimentally. For the computational basis, the three eigenstates of the system can be encoded in the frequency basis as

$$|0\rangle = |\omega_p - \Delta\rangle$$
$$|1\rangle = |\omega_p\rangle , \qquad (4)$$
$$|2\rangle = |\omega_p + \Delta\rangle$$

where $\omega_p$ is the central frequency of the narrowband photon, and $\Delta$ the frequency shift due to EO modulation. A frequency shift of a photon corresponds to a linear phase modulation in time, and the phase induced by the EOM is $e^{\pm i\Delta t}$ for negative and positive frequency shifts respectively. Because the phase is defined modulo $2\pi$, the EOM can be driven by a sawtooth voltage signal with the same slope in the ramp as the ideal linear phase ramp. Superpositions of the computational basis states (Eq. 4) are obtained by driving the EOM with different waveforms, which generally introduce photon loss. Restricting ourselves to phase modulation, we can minimize such loss. In particular, we studied the case of a specific set of sinusoidal phase modulations, with the state after modulation given by

$$|S_\theta\rangle \approx J_{-1}(\beta,\theta)|\omega_p - \Delta\rangle + J_0(\beta,\theta)|\omega_p\rangle + J_1(\beta,\theta)|\omega_p + \Delta\rangle, \qquad (5)$$

where $J_i$ is the $i$-th order Bessel function of the first kind, $\beta$ is the modulation index and $\theta$ is the modulation angle. We have kept up to the first order in Eq. 5 and omitted higher-order contributions. We wish to find values of $\beta$ and $\theta$ so that the two states given by

$$|S_+\rangle \approx J_{-1}(\beta_+,\theta_+)|\omega_p - \Delta\rangle + J_0(\beta_+,\theta_+)|\omega_p\rangle + J_1(\beta_+,\theta_+)|\omega_p + \Delta\rangle$$
$$|S_-\rangle \approx J_{-1}(\beta_-,\theta_-)|\omega_p - \Delta\rangle + J_0(\beta_-,\theta_-)|\omega_p\rangle + J_1(\beta_-,\theta_-)|\omega_p + \Delta\rangle \qquad (6)$$

are practically orthogonal, meaning that if the photon is prepared in state $|S_+\rangle$ ($|S_-\rangle$) and measured in $|S_-\rangle$ ($|S_+\rangle$), no photons will reach the detectors. A natural choice would be $\Delta\theta = |\theta_+ - \theta_-| = \pi$ and $\beta_+ = \beta_- = 1.2024$, which is half of the carrier-suppressed modulation index. The value of $\beta$ is chosen so that modulation for one symbol and demodulation for the other combine to fully suppress the carrier, which is the frequency component detected by our receiver in the end. $|S_+\rangle$ and $|S_-\rangle$ will be used as the superposition basis vectors in what follows. Without loss of generality, we omit the third orthogonal basis vector for simplicity.

**Table 1** Phase profiles written on EOM for state preparation and projection. The five profiles shown are represented as phasors of amplitude 1, which form a subset of all possible profiles that can be achieved by amplitude and phase modulation.

| State | Preparation | Projection |
|---|---|---|
| $\|0\rangle$ | $e^{i\Delta t}$ | $e^{-i\Delta t}$ |
| $\|1\rangle$ | 1 | 1 |
| $\|2\rangle$ | $e^{-i\Delta t}$ | $e^{i\Delta t}$ |
| $\|S_+\rangle \approx 0.50\|0\rangle + 0.67\|1\rangle - 0.50\|2\rangle$ | $e^{i(1+\beta\sin(\Delta t))}$ | $e^{i(1-\beta\sin(\Delta t))}$ |
| $\|S_+\rangle \approx -0.50\|0\rangle + 0.67\|1\rangle + 0.50\|2\rangle$ | $e^{i(1-\beta\sin(\Delta t))}$ | $e^{i(1+\beta\sin(\Delta t))}$ |

The state preparation of the photon is completed with one EOM, and the state projection in the receiver is performed with another EOM and a filter cavity – the detection cavity. There is a trade-off in the detection cavity bandwidth between error rate and detection efficiency. The optimal bandwidth for maximum information transfer was found to be roughly $1.5\Delta_{photon}$, with $\Delta_{photon}$ the single photon bandwidth[18]. For easier alignment, this cavity is designed to have a confocal geometry (Fig.1), that is, the separation of the cavity mirrors is set to be the same as their radius of curvature: $R_1 = R_2 = L = 100$ mm. The free spectral range of this cavity is thus $FSR_{det} = 1.5$ GHz, which is much broader than the frequency range of the EOM encoding. The measured linewidth of the cavity is $7.0 \pm 0.4$ MHz, which agrees qualitatively with the linewidth estimation from the $FSR_{det}$ and the specified reflectivity of the mirrors (99%). An arbitrary waveform generator is used to provide phase-locked driving signals of EOMs for both the modulation and demodulation, and the resulting spectra were first characterized with classical light. As shown in Table 1, the preparation and projection phase profiles are given for the five quantum states $\{|0\rangle, |1\rangle, |2\rangle, |S_+\rangle, |S_-\rangle\}$. For the initial characterization, the photons are prepared in state $|1\rangle$ and projected to each of the five states. The frequency spectrum (as a function of frequency shifts to the carrier frequency) are then recorded with a photodiode as shown in Fig. 2. We then performed a truth table measurement, yielding the pair-wise overlap of the five quantum states (see Table 2 and Fig. 3).

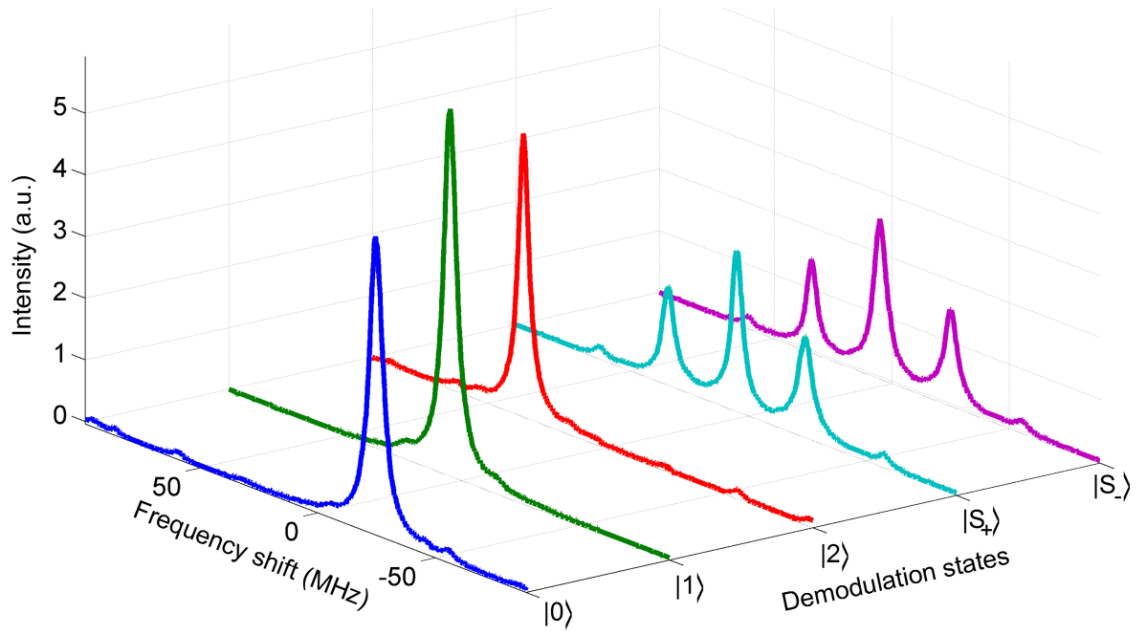

**Figure 2. Characterization of the measurement system using classical pulses.** The frequency spectrum (as a function of frequency shifts to the carrier frequency) after transmitting a $|0\rangle$ symbol and having the receiver carry out measurements for each of the five states, as discussed in the text.

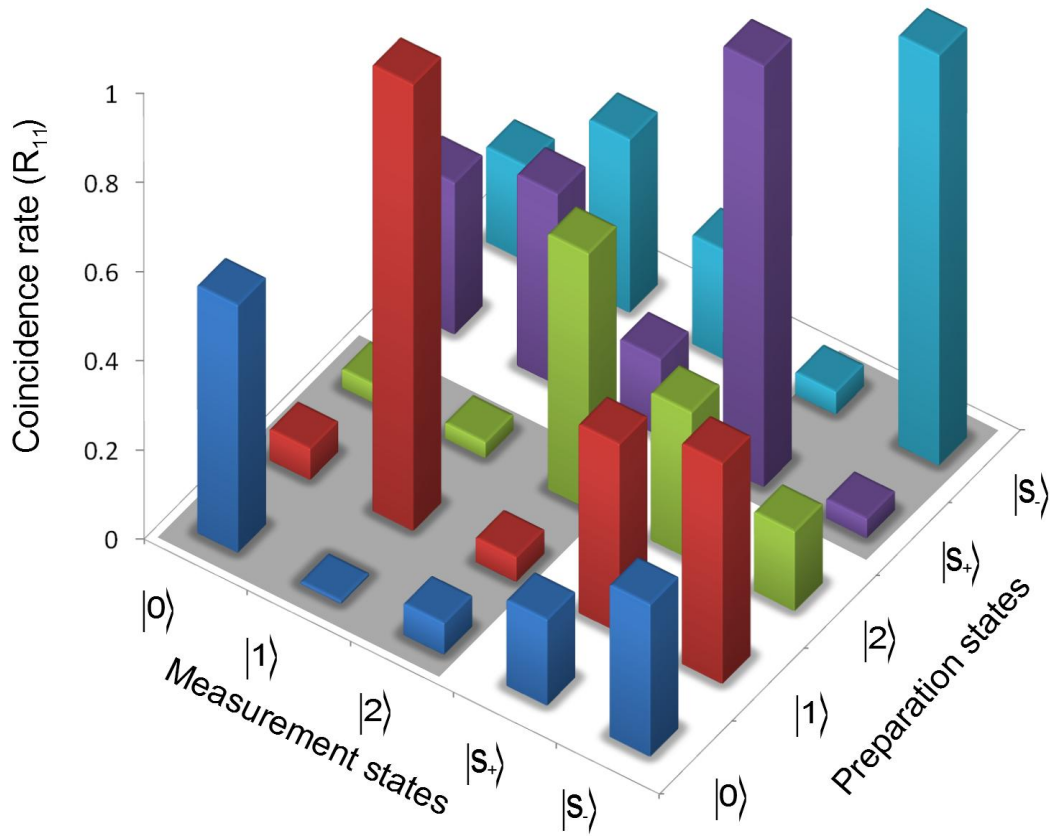

**Figure 3. State projection probability measurements.** Visualization of the relative coincidence counts. The variation of height across the diagonal is from losses due to the bandwidth limitation of the electronics. The near-zero height of coincidence counts between $|S_+\rangle$ and $|S_-\rangle$ basis shows good phase coherence of the encoding. Two ($3\times 3$ and $2\times 2$) diagonal blocks can be seen clearly for the vectors drawn from the two different bases.

**Table 2** Coincidence rates between the trigger photon and encoded photon for each combination of encoding and decoding basis. All the count rates are relative to $R_{11}$, the rate of encoding and decoding in state $|1\rangle$.

|         | $|0\rangle$ | $|1\rangle$ | $|2\rangle$ | $|S_+\rangle$ | $|S_-\rangle$ |
|---------|-------------|-------------|-------------|---------------|---------------|
| $|0\rangle$ | 0.55±0.02 | 0.005±0.002 | 0.069±0.008 | 0.19±0.01 | 0.34±0.02 |
| $|1\rangle$ | 0.068±0.008 | 1.00±0.03 | 0.054±0.007 | 0.42±0.02 | 0.49±0.02 |
| $|2\rangle$ | 0.040±0.006 | 0.034±0.005 | 0.57±0.02 | 0.33±0.02 | 0.18±0.01 |
| $|S_+\rangle$ | 0.34±0.02 | 0.43±0.02 | 0.17±0.01 | 0.94±0.03 | 0.042±0.006 |
| $|S_-\rangle$ | 0.21±0.01 | 0.39±0.02 | 0.24±0.01 | 0.049±0.006 | 0.92±0.03 |

In the two ($3\times3$ and $2\times2$) diagonal blocks of the measurement matrix (Table 2 and Fig.3), where the measurements are done within the same basis, the off-diagonal elements in the measurement matrix (*i.e.* rates when the encoding and decoding profiles are mismatched) are much lower than the diagonal elements (*i.e.* rates when encoding and decoding profiles are correctly matched). The probability should ideally be zero for the off-diagonal elements in the measurement matrix; however, there is a finite probability of cross-talk induced detection, which is measured to be less than 12.3% in total. We attribute the finite cross-talk for different vectors in the same basis to the finite bandwidth of EOM driving electronics, EOM polarization alignment, single-photon state purity and cavity fluctuation. Furthermore, due to finite bandwidth in the electronics, the photons prepared in state $|0\rangle$ and $|2\rangle$ get extra spectral components, resulting in additional losses in these states, thus the variation in the diagonal elements of the measurement matrix. We verify that the coherence between computational basis states is preserved, by confirming the orthogonality between the superposition states $|S_+\rangle$ and $|S_-\rangle$. To further investigate the coherent behavior of the superposition vectors, we varied the phase of the state preparation (Eq. 5), and then projected the photon onto the $|S_+\rangle$ state. The result is shown in Fig. 4 where one can clearly see the signature of the coherence when the phase of the state is varied, in good agreement with the quantum optical calculations, performed with no free parameters (Fig. 4). Such coherence cannot be explained by a classical model with probabilistic preparation of the different states.

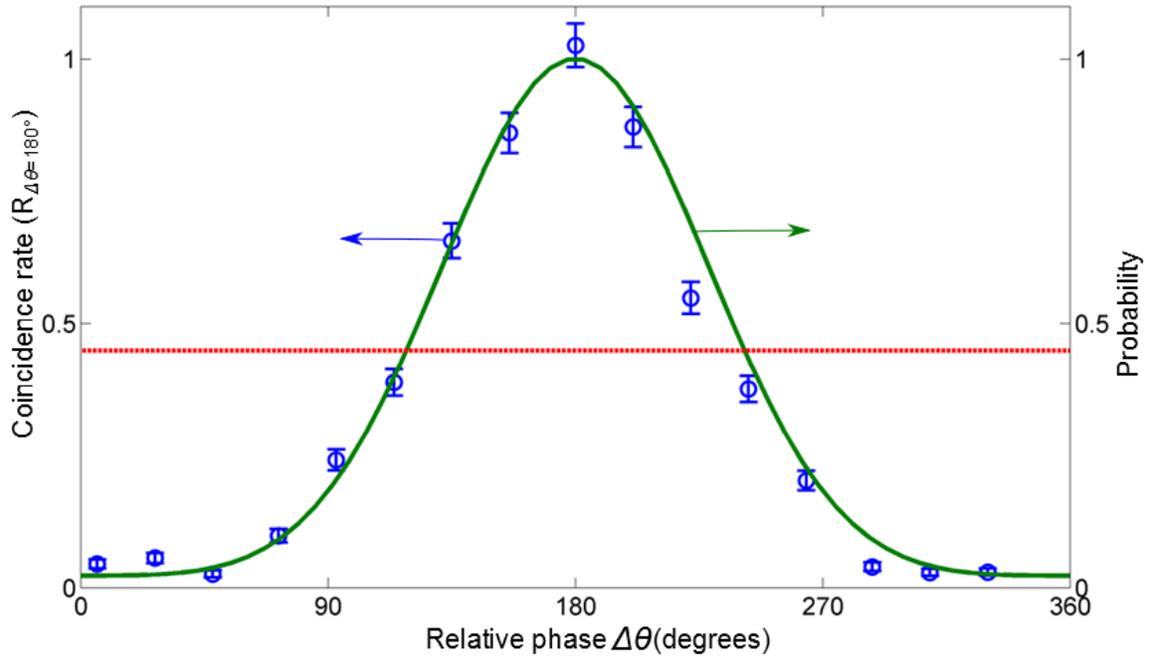

**Figure 4. Measurement and simulation of the coherence in the superposition state.** Coincidence rate (left axis) as a function of phase difference between transmitter and receiver EOMs, relative to the peak value $R_{\Delta\theta=180°}$ with its axis to the left, and predicted projection probabilities (right axis, no free parameters). The red dashed line indicates the classical detection probability of incoherent mixture of the computational basis vectors $|0\rangle$, $|1\rangle$ and $|2\rangle$ with the same weight as in $|S_+\rangle$ and $|S_-\rangle$.

In conclusion, we have experimentally demonstrated fiber-optical multidimensional quantum channel based on temporal manipulation of single photons. Our controllable-dimension temporal encoding scheme paves the way to incorporate multi-dimensional quantum information into the existing fiber optical telecommunication infrastructure.